\documentstyle[editedvolume,psfig]{crckapb}

\def\cl{\centerline}
\def\ni{\noindent}

\def\bs{\bigskip}

\def\ea{\ et al. \,}
\def\eg{{\it e.g. \,}}
\def\be{\begin{equation}}
\def\ee{\end{equation}}
\def\rel{relativistic\,}
\def\nrel{nonrelativistic\,}
\def\sz{Sunyaev \& Zeldovich\,}

\begin{opening}
\title{THE SUNYAEV-ZELDOVICH EFFECT AND \protect\\
ITS COSMOLOGICAL SIGNIFICANCE}

\author{YOEL REPHAELI}
\institute{School of Physics \& Astronomy\\
Tel Aviv University, Tel Aviv, Israel}
\end{opening} 

\runningtitle{THE S-Z EFFECT}

\begin{document}

\begin{abstract}

Comptonization of the cosmic microwave background (CMB) radiation by
hot gas in clusters of galaxies - the Sunyaev-Zeldovich (S-Z) effect -
is of great astrophysical and cosmological significance. In recent years
observations of the effect have improved tremendously; high
signal-to-noise images of the effect (at low microwave frequencies)
can now be obtained by ground-based interferometric arrays. In the
near future, high frequency measurements of the effect will be made
with bolomateric arrays during long duration balloon
flights. Towards the end of the decade the PLANCK satellite will 
extensive S-Z surveys over a wide frequency range. Along with
the improved observational capabilities, the theoretical description
of the effect and its more precise use as a probe have been
considerably advanced. I review the current status of theoretical 
and observational work on the effect, and the main results from its 
use as a cosmological probe.
\end{abstract}

\section{Introduction}

The cosmological significance of the spectral signature imprinted on the 
CMB by Compton scattering of the radiation by electrons in a hot 
intergalactic medium was realized early on by Zeldovich \& Sunyaev 
(1969). A more important manifestation of this effect occurs in clusters 
of galaxies (Sunyaev \& Zeldovich 1972). Measurements of this S-Z effect 
with single-dish radio telescopes were finally successful some 15 years 
later (for general reviews, see Rephaeli 1995a, Birkinshaw 1999). Growing 
realization of the cosmological significance of the effect has led to 
major improvements in observational techniques, and to extensive 
theoretical investigations of its many facets. The use of interferometric 
arrays, and the substantial progress in the development of sensitive 
radio receivers, have led to first images of the effect (Jones 1993, 
Carlstrom \ea 1996). Some 40 cluster images have already been obtained 
with the OVRO and BIMA arrays (Carlstrom \ea 1999, Carlstrom \ea 2001). 
Theoretical treatment of the S-Z effect has also improved, starting with 
the work of Rephaeli (1995b), who performed an exact relativistic 
calculation and demonstrated the need for such a more accurate 
description. 

The S-Z effect is essentially independent of the cluster redshift, a 
fact that makes it unique among cosmological probes. Measurements of the
effect yield directly the properties of the hot intracluster (IC)
gas, and the {\it total} dynamical mass of the cluster, as well as
indirect information on the evolution of clusters. Of more basic
importance is the ability to determine the Hubble ($H_0$) constant
and the density parameter, $\Omega$, from S-Z and X-ray measurements. 
This method to determine $H_0$, which has clear advantages over the 
traditional galactic distance ladder method based on optical 
observations of galaxies in the nearby universe, has been yielding 
increasingly higher quality results. But substantial systematic 
uncertainties, due largely to modeling of the thermal and spatial 
distributions of IC gas, need still to be reduced in order to fully 
exploit the full potential of this method. Sensitive spectral and 
spatial mapping of the effect, and the minimization of systematic 
uncertainties in the S-Z and X-ray measurements, constitute the main 
challenges of current and near future work in this rapidly progressing 
research area.

I describe the effect and review some of the recent observational
and theoretical work that significantly improved its use as a precise
cosmological probe. Progress anticipated in the near future is briefly
discussed.

\section{Exact Description of the Effect}

The scattering of the CMB by hot gas heats the radiation, resulting 
in a systematic transfer of photons from the Rayleigh-Jeans (R-J) to the 
Wien side of the (Planckian) spectrum. An accurate description 
of the interaction of the radiation with a hot electron gas necessitates 
the calculation of the exact frequency re-distribution function in the 
context of a relativistic formulation. The calculations of Sunyaev \& 
Zeldovich (1972) are based on a solution to the Kompaneets (1957) 
equation, a {\it \nrel} diffusion approximation to the exact kinetic 
(Boltzmann) equation describing the scattering. The result of their 
treatment is a simple expression for the intensity change resulting from 
scattering of the CMB (temperature $T$) by electrons with {\it thermal} 
velocity distribution (temperature $T_e$), 
\be
\Delta I_{t} = i_{o} y g(x) \; ,
\ee
where $i_{o} = 2(kT)^3 /(hc)^2$. The spatial dependence is contained in 
the Comptonization parameter, $y = \int (kT_e/mc^2) n \sigma _T dl$,
a line of sight integral (through the cluster) over the electron density 
($n$); $\sigma _T$ is the Thomson cross section. 
The spectral function, 
\be
g(x)={x^4e^x \over (e^x -1)^2} \left[{x (e^x +1)\over e^x -1}-4 \right] ,
\ee
where $x\equiv h\nu/kT$ is the non-dimensional frequency, is negative in 
the R-J region and positive at frequencies above a critical value, 
$x=3.83$, corresponding to $\sim 217$ GHz. Typically in a rich cluster 
$y \sim 10^{-4}$ along a line of sight through the center, and the 
magnitude of the relative temperature change due to the thermal effect 
is $\Delta T_t/T = -2y$ in the R-J region.  

The above thermal intensity change is the full effect only if the cluster 
is at rest in the CMB frame. Generally, the effect has a second component 
when the cluster has a finite (peculiar) velocity in the CMB frame. This 
{\it kinematic} (Doppler) component is 
\be
\Delta I_k = {x^{4}e^{x}\over (e^x -1)^2} {v_r \over c} \tau \; ,
\ee
where $v_r$ is the line of sight component of the cluster peculiar 
velocity, $\tau$ is the Thomson optical depth of the cluster. The 
related temperature change is $\Delta T_k/T = -(v_r/c)\tau$ (Sunyaev 
\& Zeldovich 1980). 

The quantitative \nrel description of the two components of the S-Z 
effect by Sunyaev \& Zeldovich (1972) is generally valid at low gas 
temperatures and at low frequencies. Rephaeli (1995b) has shown that 
this approximation is insufficiently accurate for use of the effect as 
a precise cosmological probe: Electron velocities in the IC gas are 
high, and the relative photon energy change in the scattering is 
sufficiently large to require a \rel calculation. Using the exact 
probability distribution in Compton scattering, and the relativistically 
correct form of the electron Maxwellian velocity distribution, Rephaeli 
(1995b) calculated $\Delta I_t$ in the limit of small $\tau$, keeping 
terms linear in $\tau$ (single scattering). Results of this 
semi-analytic calculation, shown in Figure 1, demonstrate that the 
relativistic spectral distribution of the intensity change is quite 
different from that derived by Sunyaev \& Zeldovich (1972). Deviations 
from their expression increase with $T_e$ and can be quite substantial. 
These are especially large near the crossover frequency, which shifts to 
higher values with increasing gas temperature. 

\begin{figure}[t]
\cl{\psfig{file=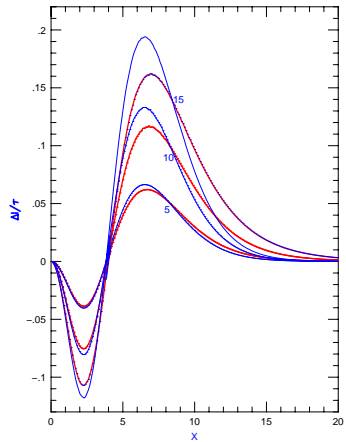,width=7cm,angle=270}}
\bs
\bs
\caption{The spectral distribution of $\Delta I_t /\tau$ (in units of 
$h^{2}c^{2}/2k^{3}T^{3}$). The pairs of thick and thin lines, 
labeled with $kT_e =$ 5, 10, and 15 keV, show the \rel and 
nonrelativistic distributions, respectively.} 
\end{figure}

The results of the semi-analytic calculations of Rephaeli (1995b) 
generated considerable interest which led to various generalizations and 
extensions of the relativistic treatment. Challinor \& Lasenby (1998) 
generalized the Kompaneets equation and obtained analytic approximations 
to its solution for the change of the photon occupation number by means 
of a power series in $\theta_{e}$ = $kT_{e}/mc^{2}$. Itoh \ea (1998) 
adopted this approach and improved the accuracy of the analytic 
approximation by expanding to fifth order in $\theta_{e}$. Sazonov \& 
Sunyaev (1998) and Nozawa \ea (1998) have extended the relativistic 
treatment also to the kinematic component obtaining -- for the first 
time -- the leading cross terms in the expression for the total 
intensity change ($\Delta I_t + \Delta I_k$) which depends on both $T_e$ 
and $v_r$. An improved analytic fit to the numerical solution, valid for 
$0.02 \leq \theta_{e} \leq 0.05$, and $x \leq 20$ ($\nu \leq 1130$ GHz), 
has recently been given by Nozawa \ea (2000). In view of the possibility 
that in some rich clusters $\tau \sim 0.02-0.03$, the approximate 
analytic expansion to fifth order in $\theta_{e}$ necessitates also the 
inclusion of multiple scatterings, of order $\tau^2$. This has been 
accomplished by Itoh \ea (2000), and Shimon \& Rephaeli (2001).

The more exact \rel description of the effect should be used in all 
high frequency S-Z work, especially when measurements of the effect 
are used to determine precise values of the cosmological parameters. 
Also, since the ability to determine peculiar velocities of clusters 
depends very much on measurements at (or very close to) the crossover 
frequency, its exact value has to be known. This necessitates knowledge 
of the gas temperature since in the exact \rel treatment the crossover 
frequency is no longer independent of the gas temperature, and is 
approximately given by 
$\simeq 217[1+1.167 kT_e/mc^2 -0.853(kT_e/mc^2)^2]$ 
GHz (Nozawa \ea 1998a). Use of the S-Z effect as a cosmological probe 
necessitates also X-ray measurements to determine (at least) the gas 
temperature. Therefore, a relativistically correct expression for the 
(spectral) bremsstrahlung emissivity must be used (Rephaeli \& Yankovitch 
1997). In the latter paper first order relativistic corrections to the 
velocity distribution, and electron-electron bremsstrahlung, were taken 
into account in correcting values of $H_0$ that were previously derived 
using the \nrel expression for the emissivity (see also Hughes \& 
Birkinshaw 1998). Nozawa \ea (1998b) have performed a more exact 
calculation of the \rel bremsstrahlung Gaunt factor.

Scattering of the CMB in clusters affects also its polarization towards 
the cluster. Net polarization is induced due to the quadrupole component 
in the spatial distribution of the radiation, and when the cluster 
peculiar velocity has a component transverse to the line of sight, 
$v_{\perp}$ (\sz 1980, Sazonov \& Sunyaev 1999). The leading 
contributions to the latter, kinematically-induced polarization, are 
proportional to $(v_{\perp}/c) \tau^2$ and $(v_{\perp}/c)^2\tau$. 
Itoh \ea (2000) have included relativistic corrections in the expression 
they derived for the kinematically induced polarization.

\section{Recent Measurements}

The quality of S-Z measurements has increased very significantly over the 
last seven years mainly as a result of the use of interferometric arrays. 
Telescope arrays have several major advantages over a single dish, 
including insensitivity of the measurements to changes in the 
atmospheric emission, sensitivity to specific angular scales and to 
signals which are correlated 
between the array elements, and the high angular resolution that enables 
nearly optimal subtraction of signals from point sources. With the 
improved sensitivity of radio receivers it became feasible to use 
interferometric arrays for S-Z imaging measurements (starting with the 
use of the Ryle telescope by Jones \ea 1993). Current state-of-the-art 
work is done with the BIMA and OVRO arrays; images of some 35 moderately 
distant  clusters (in the redshift range $0.17 - 0.89$) have already 
been obtained at $\sim 30$ GHz (Carlstrom \ea 1999, 2001). 
The S-Z image of the cluster CL0016+16, observed at 28 GHz with the BIMA 
array (Carlstrom \ea 1999), is shown in the upper frame in Figure 2. The 
ROSAT X-ray image is superposed on the contour lines of the S-Z profile 
in the lower frame. These images nicely demonstrate the good agreement 
between the orientation of the X-ray and S-Z brightness distributions, as 
well as the relative smallness of the X-ray size ($\propto n^2$) in 
comparison with the S-Z size ($\propto n$) of the cluster.
\begin{figure}
\cl{\psfig{file=fig2_1.epsi,width=5cm,angle=270}}
\cl{\hspace{0.25in}{\psfig{file=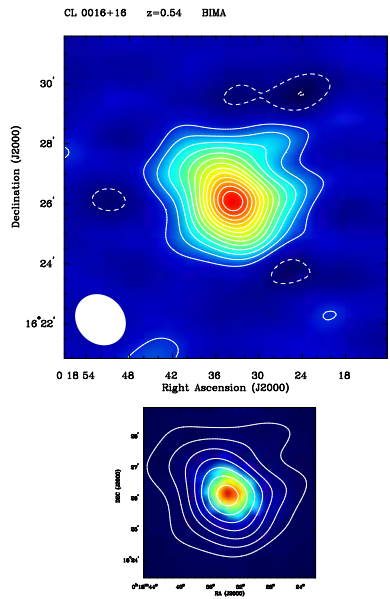,width=2.9cm,angle=0}}}
\caption{\small{S-Z and X-ray images of the cluster CL0016+16. The S-Z
image of the cluster, obtained with the BIMA array, is shown
in the upper frame. In the lower frame, contours of the S-Z effect in the
cluster are superposed on the {\it ROSAT} X-ray image (from
Carlstrom \ea 1999).}}
\end{figure}  
Work has begun recently with the CBI, a new radio (26-36 GHz) 
interferometric array (in the Chilean Andes) of small (0.9m) dishes on a 
platform with baselines in the 1m to 6m range. The spatial resolution of 
the CBI is in the $3'-10'$ range, so (unlike the BIMA and OVRO arrays) it 
is suitable for S-Z measurements of nearby clusters. Some 9 clusters have 
already been observed with the CBI (Udomprasert \ea 2000).
        
S-Z observations at higher frequencies include measurements with the 
SuZIE array, and the PRONAOS and MITO telescopes. Three moderately 
distant clusters were measured with the small (2X3) SuZIE array: A1689 
\& A2163 (Holzapfel \ea 1997a, 1997b), and recently the cluster A1835 
was observed at three spectral bands centered on 145, 221, 279 GHz 
(Mauskopf \ea 2000). Observations at four wide spectral bands (in the 
overall range of 285-1765 GHz) were made of A2163 with the PRONAOS 
atmospheric 2m telescope (Lamarre \ea 1998), leading to what seems to 
be the first detection of the effect by a balloon-borne experiment. The 
new 2.6m MITO telescope -- which currently operates at four spectral 
bands and has a large $\sim 17'$ beam -- was used to observe the effect 
in the Coma cluster (D'Alba \ea 2001).

The sample of observed clusters now includes the distant (z=0.45) cluster 
RXJ 1347 which was measured to have the largest determined Comptonization 
parameter, $y = 1.2\times 10^{-3}$ (Pointecouteau \ea 1999). The 
observations were made with the Diabolo bolometer operating at the IRAM 
30m radio telescope. The Diabolo has a $0.5'$ beam, and a dual channel 
bolometer (centered on 2.1 and 1.2 mm). Four other clusters were also 
observed with the Diabolo bolometer (Desert \ea 1998).

\section{The S-Z Effect as a Cosmological Probe}

The fact that the S-Z effect is (essentially) independent of the cluster 
redshift makes it a unique cosmological probe. This feature, its simple 
well understood nature, and the expectation that the inherent systematic 
uncertainties associated mainly with modeling of the IC gas can be 
reduced, have led to ever growing observational and theoretical interest 
in using the effect for the determination of cluster properties and 
global cosmological parameters. (For extensive discussions of use of the 
effect as a probe, see the reviews by Rephaeli 1995a, and Birkinshaw 
1999.) 

In principle, high spatial resolution measurements of the S-Z effect 
yield the gas temperature and density profiles across the cluster. 
Cluster gas density and temperature profiles have so far been mostly 
deduced from X-ray measurements. S-Z measurements can more fully 
determine these profiles due to the linear dependence of $\Delta I_t$ on 
$n$ (and $T$), as compared to the $n^2$ 
dependence of the (thermal bremsstrahlung) X-ray brightness profile. 
This capability has been reached for the first time in the analysis of 
the interferometric BIMA and OVRO images (Carlstrom \ea 2001).

The cluster full mass profile, $M(r)$, can be derived directly from the 
gas density and temperature distributions by solving the equation of 
hydrostatic equilibrium (assuming, of course, the gas has reached such a 
state in the underlying gravitational potential). This method has already 
been employed in many analyses using X-ray deduced gas parameters (\eg 
Fabricant \ea 1980). Grego \ea (2001) have recently used this method 
to determine total masses and gas mass fractions of 18 clusters based 
largely on the results of their interferometric S-Z measurements. 
Isothermal gas with the familiar density profile, 
$(1 + r^2/r_c^2)^{-3\beta/2}$, was assumed. The core radius, $r_c$, and 
$\beta$ were determined from analysis of the S-Z data, whereas the X-ray 
value of the temperature was adopted. Mean values of the gas mass 
fraction were found to be in the range $(0.06-0.09)h^{-1}$ (where $h$ 
is the value of $H_0$ in units of $100$ km s$^{-1}$ Mpc$^{-1}$) for the 
currently popular open and flat, $\Lambda$-dominated, CDM models.
Note that the deduction of an approximate 3D density and temperature 
distributions from their sky-projected profiles requires use of a 
deprojection algorithm. Such an algorithm was developed by Zaroubi 
\ea (1998).

Measurement of the kinematic S-Z effect yields the line of sight 
component of the cluster peculiar velocity ($v_r$). This is 
observationally feasible only in a narrow spectral band near the
critical frequency, where the thermal effect vanishes while the kinematic
effect -- which is usually swamped by the much larger thermal component --
is maximal (Rephaeli \& Lahav 1991). SuZIE is the first experiment with 
a spectral band centered on the crossover frequency. Measurements of 
the clusters A1689 and A2163 (Holzapfel \ea 1997b) and A1835 (Mauskopf 
\ea 2000) yielded substantially uncertain results for $v_r$ 
($170^{+815}_{-630}$, $490^{+1370}_{-880}$, and $500 \pm 1000$ 
km s$^{-1}$, respectively). Balloon-borne measurements of the effect with 
PRONAOS have also yielded insignificant determination of the peculiar 
velocity of A2163 (Lamarre \ea 1998).

Most attention so far has been given to the determination of the Hubble 
constant, $H_0$, and the cosmological density parameter, $\Omega$, from 
S-Z and X-ray measurements. Briefly, the method is based on determining 
the angular diameter distance, $d_A$, from a combination of 
$\Delta I_t$, the X-ray surface brightness, and their spatial profiles. 
Averaging over the first eight determinations of $H_0$ (from S-Z and 
X-ray measurements of seven clusters) yielded $H_0 \simeq 58 \pm 6$ km 
s$^{-1}$ Mpc$^{-1}$ (Rephaeli 1995a), but the database was then very 
non-uniform and the errors did not include systematic uncertainties. 
Repeating this with a somewhat updated data set, Birkinshaw (1999) 
deduced essentially a similar mean value ($60$ km s$^{-1}$ Mpc$^{-1}$), 
but noted that the individual measurements are not independent, so that 
a simple error estimation is not very meaningful. A much larger S-Z data 
set is now available from the interferometric BIMA and OVRO 
observations, and since the redshift range of the clusters in the sample 
is substantial, the dependence on $\Omega$ is appreciable. A fit to 33 
cluster distances gives $H_0 = 60$ km s$^{-1}$ Mpc$^{-1}$ for 
$\Omega = 0.3$, and $H_0 = 58$ km s$^{-1}$ Mpc$^{-1}$ for $\Omega = 1$, 
with direct observational errors of $\pm 5\%$ (Carlstrom \ea 2001). The 
main known sources of systematic uncertainties (see discussions in 
Rephaeli 1995a, and Birkinshaw 1999) are presumed to introduce an 
additional error of $\sim 30\%$ (Carlstrom \ea 2001). The current number 
of clusters with S-Z determined distances is sufficiently large that a 
plot of $d_A$ vs. redshift (a Hubble diagram) is now quite of interest, 
but with the present level of uncertainties the limits on the value of 
$\Omega$ are not very meaningful (as can be seen from figure 11 of 
Carlstrom \ea 2001).

The S-Z effect induces anisotropy in the spatial distribution of the CMB 
(Sunyaev 1977, Rephaeli 1981); this is the main source of secondary 
anisotropy on angular scales of few arcminutes. The magnitude of the 
temperature anisotropy, $\Delta T/T$ can be as high as $few \times 
10^{-6}$ on angular scales of a few arcminutes, if the gas evolution in 
clusters is not too strong (Colafrancesco \ea 1994). Because of this, 
and the considerable interest in this range of angular scales -- 
multipoles (in the representation of the CMB temperature in terms of 
spherical harmonics) $\ell \geq 1000$ -- the S-Z anisotropy has been 
studied extensively in the last few years. The basic goal has been to 
determine the S-Z anisotropy in various cosmological, large scale 
structure, and IC gas models. The anisotropy is commonly characterized 
by the $\ell$ dependence of its power spectrum. The strong motivation 
for this is the need to accurately calculate the power spectrum of the 
full anisotropy in order to make precise global parameter determinations 
from the analysis of large stratospheric and satellite databases. In 
addition, mapping the S-Z anisotropy will yield direct information on 
the evolution of the cluster population.

Results from many calculations of the predicted S-Z anisotropy are not 
always consistent even for the same cosmological and large scale 
structure parameters. This is simply due to the fact that the calculation 
involves a large number of input parameters in addition to the global 
cosmological parameters (\eg the present cluster density, and parameters 
characterizing the evolutionary history of IC gas), and the 
sensitive dependence of the anisotropy on some of these. 
The anisotropy can be more directly estimated from simulations of the S-Z 
sky, based largely on results from cluster X-ray surveys and the use of 
simple scaling relations (first implemented by Markevitch \ea 1992).

The observational capabilities of upcoming long duration balloon-borne 
experiments and satellites are expected to result in detailed 
mapping of the small angular scale anisotropy. These have motivated many 
recent works; Colafrancesco \ea (1997), and Kitayma \ea (1998), have 
calculated the S-Z cluster number counts in an array of open and flat 
cosmological and dark matter models, and da Silva \ea (1999) have 
carried out hydrodynamical simulations in order to generate S-Z maps and 
power spectra. Cooray \ea (2000) have, in particular, concluded 
that the planned multi-frequency survey with the Planck satellite should 
be able to distinguish between the primary and S-Z anisotropies, and 
measure the latter with sufficient precision to determine its power 
spectrum and higher order correlations.

The main characteristics of the predicted power spectrum of the induced 
S-Z anisotropy are shown in Figure 3. Plotted are the angular power 
spectra, $C_{\ell}(\ell +1)/2\pi$, vs. the multipole, $\ell$, 
for both the primary and S-Z induced anisotropies in the (currently 
fashionable) flat cosmological model with $\Omega_{\Lambda} = 0.7$ 
(where $\Lambda$ is the cosmological constant) and with a CDM density 
parameter $\Omega_M = 0.3$. The figure is from the work of Sadeh and 
Rephaeli (2001), who studied the S-Z anisotropy in the context of 
treatment (which is an extension of the approach adopted by 
Colafrancesco \ea 1997) which is based on a Press \& Schechter cluster 
mass function, normalized by the observed X-ray luminosity function 
(see also Colafrancesco \ea 1994). IC gas was assumed to evolve in a 
simple manner consistent with the results of the EMSS survey carried 
out with the Einstein satellite. The primary anisotropy was calculated 
using the CMBFAST code of Seljak \& Zaldarriaga (1996). The solid line 
shows the primary anisotropy which dominates over the S-Z anisotropy 
for $\ell < 3000$. The S-Z power is largely due to the thermal effect; 
this rises with $\ell$ and is maximal around  $\ell \sim 1000$.
In this model, the S-Z power spectrum contributes a fraction of 
5\% (10\%) to the primary anisotropy at $\ell \simeq 1520$ ($\ell 
\simeq 1840$). It can be concluded from this (and other studies) that 
the S-Z induced anisotropy has to be taken into account, if the 
extraction of the cosmological parameters from an analysis of 
measurements of the CMB power spectrum at $\ell > 1500$ is to be precise.
\begin{figure}
\cl{\psfig{file=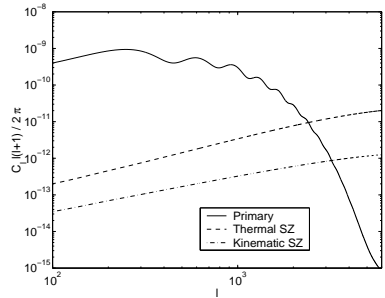,width=5cm,angle=0}}
\caption{\small{Primary and S-Z power spectra in the flat CDM model 
(Sadeh \& Rephaeli 2001). The solid line shows the primary anisotropy as 
calculated using the CMBFAST computer code of Seljak \& Zaldarriaga 
(1996). The dashed line shows the thermal S-Z power spectrum, and the 
dotted-dashed line is the contribution of the kinematic component.}}
\end{figure}   

\section{Future Prospects}

The S-Z effect is a unique cosmological and cluster probe. In the near 
future sensitive observations of the effect with ground-based and 
balloon-borne telescopes, equipped with bolometric multi-frequency 
arrays, are expected to yield high-quality measurements of its spectral 
and spatial distributions in a The capability to measure the effect at 
several high-frequency bands (in the range $150-450$ GHz) will make it 
possible to exploit the S-Z characteristic spectrum as a powerful 
diagnostic tool. The main limitation on the accuracy of the cosmological 
parameters will continue to be due to systematic uncertainties. 
Therefore, the most significant results will be obtained from 
measurements of the effect in {\it nearby} clusters, $z \leq 0.1$, where 
systematic uncertainties can be most optimally reduced. The combination 
of S-Z measurements of a large number of clusters with the new 
generation of bolometric arrays, and the improved X-ray data which are 
currently available from observations of clusters with the XMM and {\it 
Chandra} satellites, will greatly improve the accuracy of the derived 
values of cluster masses, and of the Hubble constant. In particular, it 
will be possible to measure $H_0$ with an overall uncertainty of just 
$\sim 5\%$.

\end{document}